\documentclass{moriond}

\def\be{\begin{equation}}
\def\ee{\end{equation}}
\def\bea{\begin{eqnarray}}
\def\eea{\end{eqnarray}}

\newcommand{\Photo}{\includegraphics[height=35mm]{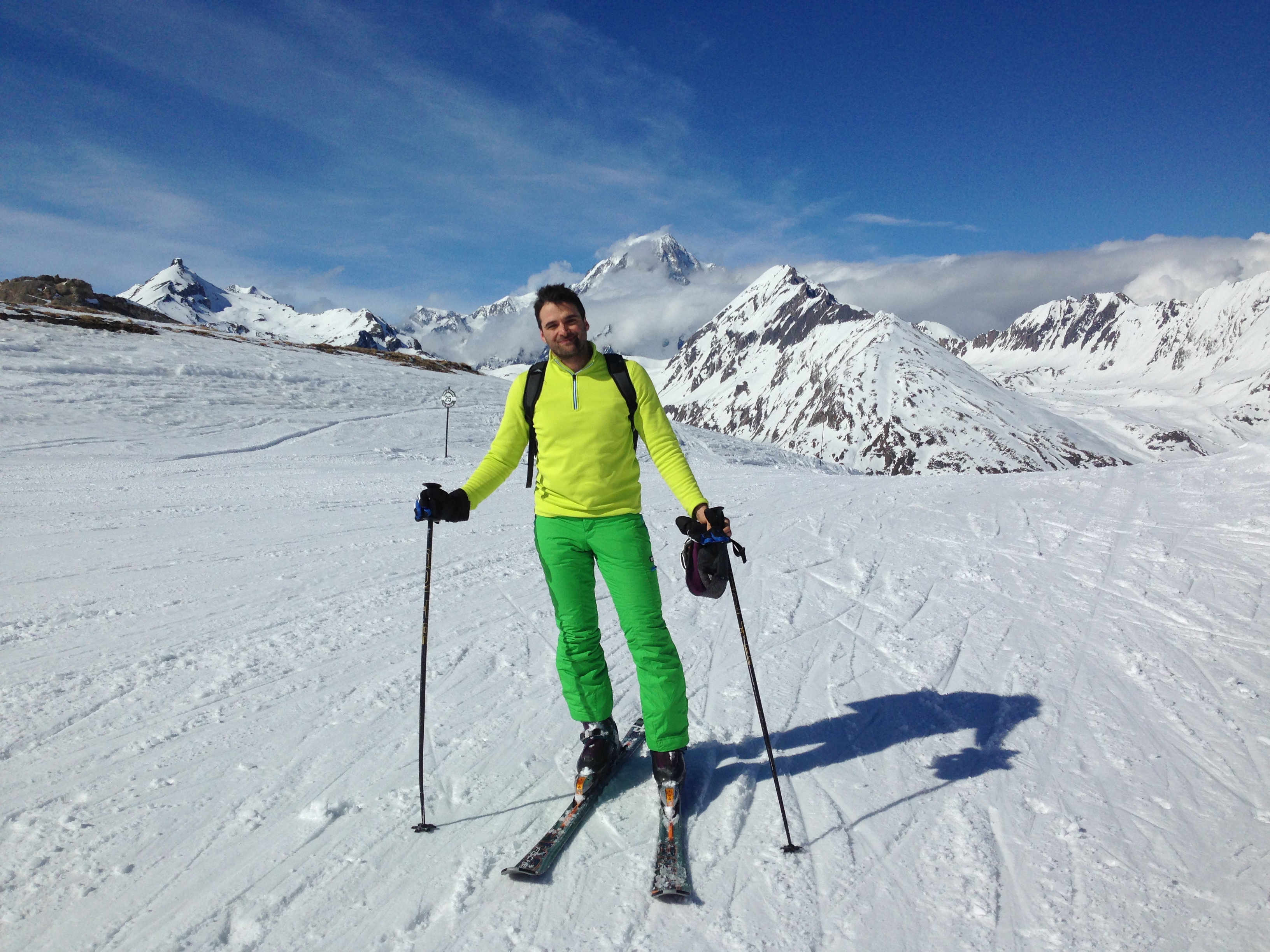}}

\begin{document}
\vspace*{4cm}
\title{A DARK MATTER SUPERFLUID}

\author{Justin Khoury}

\address{Center for Particle Cosmology, Department of Physics and Astronomy, \\ University of Pennsylvania, Philadelphia, PA 19104}

\maketitle\abstracts{In this talk we present a novel framework that unifies the stunning success of MOND on galactic scales with the triumph of the $\Lambda$CDM model
on cosmological scales. This is achieved through the rich and well-studied physics of superfluidity. The dark matter and MOND components have a
common origin, representing different phases of a single underlying substance. In galaxies, dark matter thermalizes and condenses to form a superfluid phase. The superfluid phonons couple to baryonic matter particles and mediate a MOND-like force. Our framework naturally distinguishes between galaxies (where MOND is successful) and galaxy clusters (where MOND is not): dark matter has a higher temperature in clusters, and hence is in a mixture of superfluid and normal phase. The rich and well-studied physics of superfluidity leads to a number of striking observational signatures, which we briefly discuss. Remarkably the critical temperature and equation of state of the dark matter superfluid are similar to those of known cold atom systems. Identifying a precise cold atom analogue would give important insights on the microphysical interactions underlying DM superfluidity. Tantalizingly, it might open the possibility of simulating the properties and dynamics of galaxies in laboratory experiments.}

\section{Introduction}

In the $\Lambda$-Cold-Dark-Matter ($\Lambda$CDM) standard model of cosmology, dark matter (DM) consists of collisionless particles.
This model does exquisitely well at fitting a number of large-scale observations, from the background expansion history to the 
cosmic microwave background anisotropies to the linear growth of cosmic structures~\cite{Ade:2013zuv}. 

On the scales of galaxies, however, the situation is murkier. A number of challenges have emerged for the standard $\Lambda$CDM model
in recent years, as observations and numerical simulations of galaxies have improved in tandem. For starters, galaxies in our universe are
surprisingly regular, exhibiting striking correlations among their physical properties. For instance, disc galaxies display a remarkably tight correlation between the total baryonic mass (stellar + gas) and the asymptotic rotational velocity, $M_{\rm b} \sim v_{\rm c}^4$. This scaling relation, known as the Baryonic Tully-Fisher Relation (BTFR)~\cite{McGaugh:2000sr,McGaugh:2005qe}, is unexplained in the standard model. In order to reproduce the BTFR on average, simulations must finely adjust many parameters that model complex
baryonic processes.  Given the stochastic nature of these processes, the predicted scatter around the BTFR is much larger than the observed tight correlation~\cite{Vogelsberger:2014dza}.

Another suite of puzzles comes from the distribution of dwarf satellite galaxies around the Milky Way (MW)
and Andromeda galaxies. The $\Lambda$CDM model predicts hundreds of small DM halos orbiting MW-like galaxies,
which are in principle good homes for dwarf galaxies, yet only $\sim 20-30$ dwarfs are observed around the MW and Andromeda.
Recent attempts at matching the populations of simulated subhaloes and observed MW dwarf galaxies have revealed a ``too big to fail'' problem~\cite{BoylanKolchin:2011de}: the most massive dark halos seen in the simulations are too dense to host the brightest MW satellites. Even more puzzling is the fact that the majority of the MW~\cite{Pawlowski:2012vz} and Andromeda~\cite{Ibata:2013rh,Ibata:2014pja} satellites lie within vast planar structures and are co-rotating within these planes. (Phase-space correlated dwarfs have also been found around galaxies beyond the Local Group~\cite{Ibata:2014csa}.) This suggests that dwarf satellites did not form independently, as predicted by the standard model, but may have been created through an entirely different mechanism~\cite{Pawlowski:2012vz,Zhao:2013uya}.

\begin{figure}
\begin{center}
\includegraphics[width=0.5\linewidth]{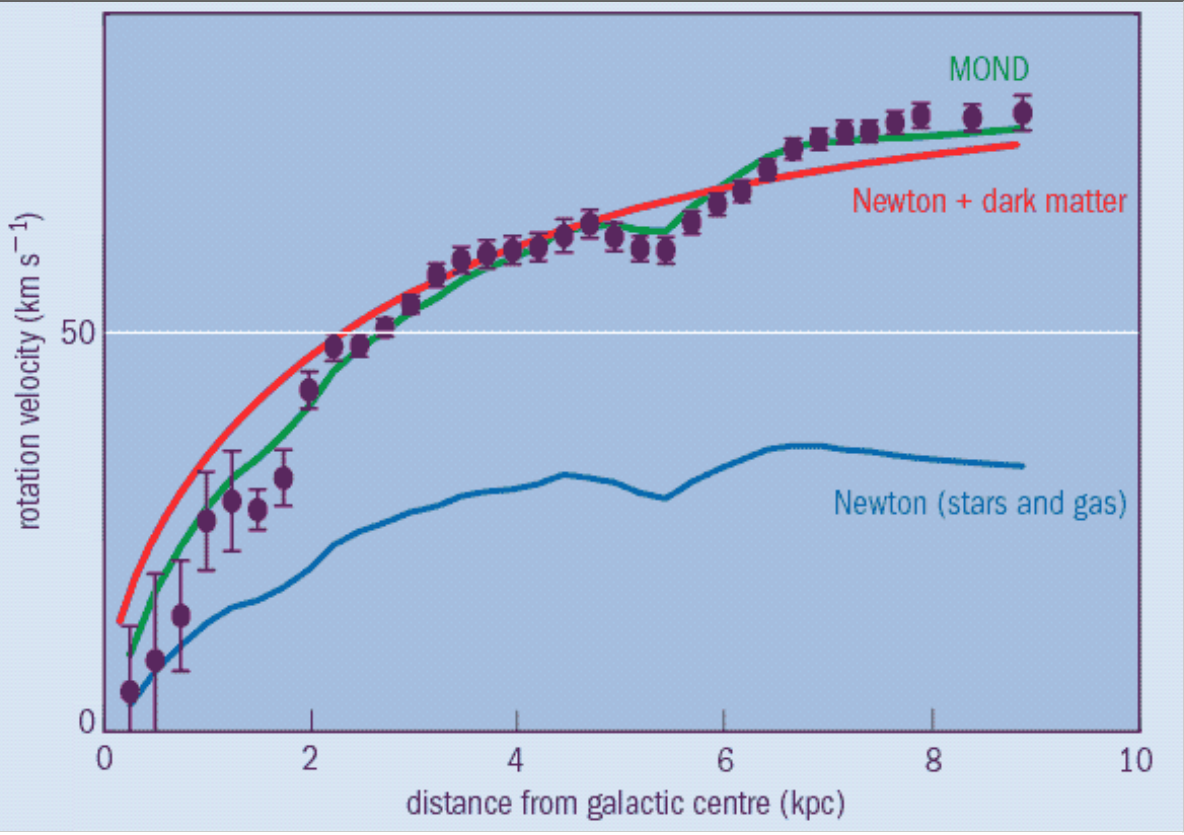}
\end{center}
\caption[]{Observed rotation curve for NGC1560 (blue points)~\cite{Broeils1992}. The MOND curve (green)~\cite{Begeman1991} offers a much better fit to the data than the $\Lambda$CDM curve (blue)~\cite{sellwood2005}. Reproduced from~\cite{Milgrom2009}.}
\label{rotcurve}
\end{figure}

A radical alternative is MOdified Newtonian Dynamics (MOND)~\cite{Milgrom:1983ca,Sanders:2002pf}. MOND replaces DM with a modification to Newton's
gravitational force law that kicks in whenever the acceleration drops below a critical value $a_0$. For large acceleration, $a\gg a_0$,
the force law recovers Newtonian gravity: $a \simeq a_{\rm N}$. At low acceleration, $a \ll a_0$, the force law is modified: $a\simeq \sqrt{a_{\rm N}a_0}$. 
This simple empirical law has been remarkably successful at explaining a wide range of galactic phenomena~\cite{Famaey:2011kh}. In particular, asymptotically flat rotation curves
and the BTFR are exact consequences of the force law.\footnote{Consider a test particle orbiting a galaxy of mass $M_{\rm b}$, in the low acceleration regime. Equating the centripetal acceleration $v^2/r$ to the MONDian acceleration $\sqrt{a_{\rm N}a_0} = \sqrt{\frac{G_{\rm N} M_{\rm b} a_0}{r^2}}$, we obtain a velocity that is independent of distance, $v^2 = \sqrt{G_{\rm N} M_{\rm b} a_0}$, in agreement with the flat rotation curves of spiral galaxies. Squaring this gives the BTFR relation $M_{\rm b} = \frac{v^4}{G_{\rm N} a_0}$
as an exact prediction.} MOND does exquisitely well at fitting detailed galactic rotation curves, as shown in Fig.~\ref{rotcurve}. There is a single parameter, 
the critical acceleration $a_0$, whose best-fit value is intriguingly of order the speed of light $c$ times the Hubble constant $H_0$: $a_0 \simeq \frac{1}{6}cH_0 \simeq 1.2\times 10^{-8}~{\rm cm}/{\rm s}^2$. 

However, the empirical success of MOND is limited to galaxies. The predicted X-ray temperature profile in massive clusters of galaxies 
is far from the observed  approximately isothermal profile~\cite{Aguirre:2001fj}. Relativistic extensions of MOND, {\it e.g.}~\cite{Bekenstein:2004ne}, fail to reproduce
CMB anisotropies and large-scale clustering of galaxies~\cite{Skordis:2005xk}. The ``Bullet" Cluster~\cite{Clowe:2003tk,Clowe:2006eq}, the aftermath of two colliding galaxy clusters, is also problematic for MOND~\cite{Angus:2006qy}.

\section{Dark Matter Condensate}

In this talk, based on two recent papers~\cite{Berezhiani:2015pia,Berezhiani:2015bqa}, we present a unified framework for the DM and MOND phenomena based on the rich and well-studied physics of superfluidity. The DM and MOND components have a common origin, representing different phases of a single underlying substance. The central idea is that DM forms a superfluid inside galaxies, with a coherence length of galactic size. 

As is familiar from liquid helium, a superfluid at finite temperature (but below the critical temperature) is best described phenomenologically as a mixture of two fluids~\cite{tisza,london,landau}: $i)$ the superfluid, which by definition has vanishing viscosity and carries no entropy; $ii)$ the ``normal" component, comprised of massive particles, which is viscous and carries entropy. The fraction of particles in the condensate decreases with increasing temperature. Thus our framework naturally distinguishes between galaxies (where MOND is successful) and galaxy clusters (where MOND is not). Galaxy clusters have a higher velocity dispersion and correspondingly higher DM temperature. For $m\sim {\rm eV}$ we will find that galaxies are almost entirely condensed, whereas galaxy clusters are either in a mixed phase or entirely in the normal phase.

As a back-of-the-envelope calculation, we can estimate the condition for the onset of superfluidity ignoring interactions among DM particles. With this simplifying
approximation, the requirement for superfluidity amounts to demanding that the de Broglie wavelength $\lambda_{\rm dB} \sim 1/mv$ of DM particles should 
be larger than the interparticle separation $\ell\sim \left(m/\rho \right)^{1/3}$. This implies an upper bound on the particle mass, $m ~\lower .75ex \hbox{$\sim$} \llap{\raise .27ex \hbox{$<$}} ~ (\rho/v^3)^{1/4}$. Substituting the value of $v$ and $\rho$ at virialization, given by standard collapse theory, this translates to~\cite{Berezhiani:2015pia,Berezhiani:2015bqa}
\begin{equation}
m ~\lower .75ex \hbox{$\sim$} \llap{\raise .27ex \hbox{$<$}} ~ 2.3 \left(1+z_{\rm vir}\right)^{3/8}\;  \left(\frac{M}{10^{12}h^{-1}M_\odot}\right)^{-1/4}\; {\rm eV}\,,
\end{equation}
where $M$ and $z_{\rm vir}$ are the mass and virialization redshift of the object. Hence light objects form a Bose-Einstein condensate (BEC) while heavy objects do not.

Another requirement for Bose-Einstein condensation is that DM thermalize within galaxies. We assume that DM particles interact through contact repulsive interactions.
Demanding that the interaction rate be larger than the galactic dynamical time places a lower bound on the interaction cross-section. For $M = 10^{12}h^{-1}M_\odot$ and
$z_{\rm vir} = 2$, the result is~\cite{Berezhiani:2015pia,Berezhiani:2015bqa}
\begin{equation}
\frac{\sigma}{m}~\lower .75ex \hbox{$\sim$} \llap{\raise .27ex \hbox{$>$}}~\left(\frac{m}{{\rm eV}}\right)^{4}\,\frac{{\rm cm}^2}{{\rm g}}\,.
\label{siglow}
\end{equation}
With $m ~\lower .75ex \hbox{$\sim$} \llap{\raise .27ex \hbox{$<$}} ~ {\rm eV}$, this is just
below the most recent constraint from galaxy cluster mergers~\cite{Harvey:2015hha}, though such
constraints should be carefully reanalyzed in the superfluid context. 

Again ignoring interactions, the critical temperature for DM superfluidity is $T_c \sim {\rm mK}$, which intriguingly is comparable to known critical temperatures for cold atom gases, {\it e.g.}, ${}^7$Li atoms have $T_c \simeq 0.2$~mK. Cold atoms might provide more than just a useful analogy --- in many ways, our DM component behaves exactly like cold atoms. In cold atom experiments, atoms are trapped using magnetic fields; in our case, DM particles are attracted in galaxies by gravity.

\section{Superfluid Phase}

Instead of behaving as individual collisionless particles, the DM is more aptly described as collective excitations: phonons and massive quasi-particles. Phonons, in particular, play a key role by mediating a long-range force between ordinary matter particles. As a result, a test particle orbiting the galaxy is subject to two forces: the (Newtonian) gravitational force and the phonon-mediated force.

Specifically, it is well-known that the effective field theory (EFT) of superfluid phonon excitations at lowest order in derivatives is a $P(X)$ theory~\cite{Son:2002zn}. Our postulate is that DM phonons are described by the non-relativistic MOND scalar action,
\begin{equation}
P(X) \sim \Lambda X\sqrt{|X|}\,;\qquad X =  \dot{\theta}  - m \Phi - \frac{(\vec{\nabla}\theta)^2}{2m}\,.
\label{PMOND}
\end{equation}
where $\Lambda\sim {\rm meV}$ to reproduce the MOND critical acceleration, and $\Phi$ is the gravitational potential. The fractional $3/2$ power would be strange if Eq.~(\ref{PMOND}) described a fundamental scalar field. As a theory of phonons, however, it is not uncommon to encounter fractional powers in cold atom systems.
For instance, the Unitary Fermi Gas (UFG)~\cite{UFGreview,Giorgini:2008zz}, which has generated much excitement recently in the cold atom community, describes a gas of cold fermionic atoms tuned at unitarity. The effective action for the UFG superfluid is uniquely fixed by 4d scale invariance at lowest-order in derivatives, ${\cal L}_{\rm UFG}(X) \sim X^{5/2}$, which is also non-analytic~\cite{Son:2005rv}.

To mediate a force between ordinary matter, $\theta$ must couple to the baryon density:
\be
{\cal L}_{\rm int} = -\alpha \frac{\Lambda}{M_{\rm Pl}} \theta \rho_{\rm b} \,,
\label{Lint}
\ee
where $\alpha$ is a dimensionless parameter. This term explicitly breaks the shift symmetry, but only at the $1/M_{\rm Pl}$ level and is therefore technically natural.
From the superfluid perspective,~Eq.~(\ref{Lint}) can arise if baryonic matter couple to the vortex sector of the superfluid, giving rise to
operators $\sim \cos\theta \rho_{\rm b}$ that preserve a discrete subgroup of the continuous shift symmetry~\cite{Villain,cosinecoupling,Randy}.

\subsection{Properties of the Condensate and Phonons}

The form of the phonon action uniquely fixes the properties of the condensate through standard thermodynamics arguments.
At finite chemical potential, $\theta = \mu t$, ignoring phonon excitations and gravitational potential to zero, the pressure of the condensate
is given as usual by the Lagrangian density,
\begin{equation}
P(\mu) = \frac{2\Lambda}{3} (2m\mu)^{3/2}\,.
\end{equation}
This is the grand canonical equation of state, $P = P(\mu)$, for the condensate. Differentiating with respect to $\mu$ yields the number density of condensed particles:
\begin{equation}
n = \frac{\partial P}{\partial\mu} = \Lambda (2m)^{3/2} \mu^{1/2}\,.
\label{nmu}
\end{equation}
Combining these expressions and using the non-relativistic relation $\rho= m n$, we find
\begin{equation}
P = \frac{\rho^3}{12\Lambda^2m^6}\,.
\label{eos}
\end{equation}
This is a polytropic equation of state $P \sim \rho^{1 + 1/n}$ with index $n = 1/2$. 

Including phonons excitations $\theta = \mu t + \phi$, the quadratic action for $\phi$ is 
\begin{equation}
{\cal L}_{\rm quad} =  \frac{\Lambda (2m)^{3/2}}{4\mu^{1/2}} \left( \dot{\phi}^2 - \frac{2\mu}{m} (\vec{\nabla}\phi)^2\right)\,.
\end{equation}
The sound speed can be immediately read off:
\begin{equation}
c_s = \sqrt{\frac{2\mu}{m}}\,.
\label{cs}
\ee

\begin{figure}
\begin{center}
\includegraphics[width=0.5\linewidth]{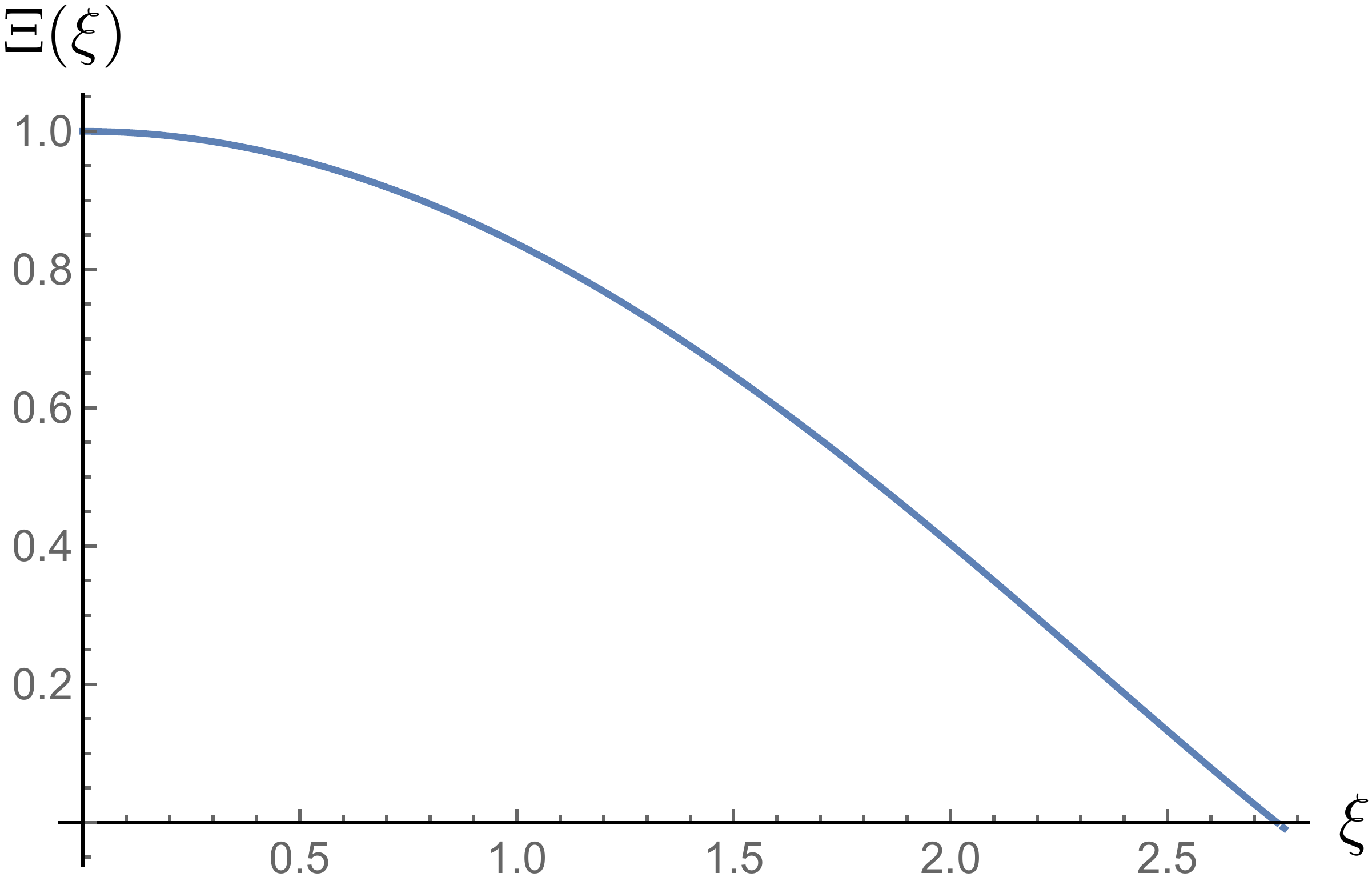}
\end{center}
\caption[]{Numerical solution of Lane-Emden equation, Eq.~(\ref{LaneEmdenEqn}).}
\label{laneemden}
\end{figure}

\subsection{Halo profile}

Assuming hydrostatic equilibrium, we can compute the density profile of a
spherically-symmetric DM condensate halo:
\begin{equation}
\frac{1}{\rho(r)}\frac{{\rm d}P(r)}{{\rm d}r} = - \frac{4\pi G_{\rm N}}{r^2} \int_0^r {\rm d}r' r'^2\rho(r')\,.
\label{hydro}
\end{equation}
Substituting the equation of state given by Eq.~(\ref{eos}), and introducing the dimensionless variables 
$\rho = \rho_0\Xi$ and $r = \sqrt{\frac{\rho_0}{32\pi G_{\rm N} \Lambda^2 m^6}}\,\xi$, with $\rho_0 $ denoting the central density, Eq.~(\ref{hydro}) implies
the Lane-Emden equation
\be
\left(\xi^2 \Xi'\right)' = - \xi^2 \Xi^{1/2}\,,
\label{LaneEmdenEqn}
\ee
where $'\equiv {\rm d}/{\rm d}\xi$. The numerical solution, with boundary conditions $\Xi(0) = 1$ and $\Xi'(0) = 0$, is shown in Fig.~\ref{laneemden}.
The superfluid density profile is cored, not surprisingly, and therefore avoids the cusp problem of CDM. 

The density is found to vanish at $\xi_1 \simeq 2.75$, which defines the halo size: $R = \sqrt{\frac{\rho_0}{32\pi G_{\rm N} \Lambda^2 m^6}}\; \xi_1$.
Meanwhile the central density is related to the halo mass as~\cite{chandrabook} $\rho_0 =  \frac{3M}{4\pi R^3} \frac{\xi_1}{|\Xi'(\xi_1)|}$,
with $\Xi'(\xi_1)\simeq -0.5$. Combining these results, it is straightforward to solve for $\rho_0$ and $R$:
\bea
\nonumber
\rho_0 &\simeq &  \left(\frac{M_{\rm DM}}{10^{12}M_\odot}\right)^{2/5} \left(\frac{m}{{\rm eV}}\right)^{18/5} \left(\frac{\Lambda}{{\rm meV}}\right)^{6/5}  \; 7 \times 10^{-25}~{\rm g}/{\rm cm}^3\,;\\ 
R &\simeq & \left(\frac{M_{\rm DM}}{10^{12}M_\odot}\right)^{1/5} \left(\frac{m}{{\rm eV}}\right)^{-6/5}\left(\frac{\Lambda}{{\rm meV}}\right)^{-2/5} \; 36~{\rm kpc}\,.
\label{halorad}
\eea
Remarkably, for $m\sim {\rm eV}$ and $\Lambda\sim {\rm meV}$ we obtain DM halos of realistic size! In the standard CDM picture a halo of mass $M_{\rm DM} = 10^{12}\,M_\odot$ has a virial radius of $\sim 200$~kpc. In our framework, the condensate radius can in principle be considerably smaller or larger depending on parameter values. For concreteness, in the remainder of the analysis we will choose as fiducial values
\be
m = 0.6~{\rm eV}\,;\qquad \Lambda = 0.2~{\rm meV}\,.
\label{fidparam}
\ee
This implies a condensate radius of $\sim 125$~kpc for a halo of mass $M_{\rm DM} = 10^{12}\,M_\odot$. 

\section{Phonon-Mediated MONDian Force}

Next we derive the phonon profile in galaxies, modeling the baryons as a static, spherically-symmetric localized source for simplicity.
We first focus on the zero-temperature analysis, where the Lagrangian is given by the sum of Eqs.~(\ref{PMOND}) and~(\ref{Lint}). 
In the static spherically-symmetric approximation, $\theta = \mu t + \phi(r)$, the equation of motion reduces~to
\be
\vec{\nabla} \cdot \left( \sqrt{2m|X|}~\vec{ \nabla}\phi\right) = \frac{\alpha\rho_{\rm b}(r)}{2M_{\rm Pl}}\,,
\ee
where $X(r) = \mu -m\Phi(r) - \frac{\phi'^2(r)}{2m}$. This can be readily integrated:
\be
\sqrt{2m|X|}~\phi' = \frac{\alpha M_{\rm b}(r)}{8\pi M_{\rm Pl} r^2}\equiv \kappa(r)\,.
\label{kappadef}
\ee
There are two branches of solutions, depending on the sign of $X$. We focus on the MOND branch (with $X < 0$):
\be
\phi' (r) = \sqrt{m}\left( \hat{\mu}  + \sqrt{\hat{\mu}^2 + \kappa^2/m^2}\right)^{1/2} \,,
\label{phigensoln}
\ee
where $\hat{\mu} \equiv \mu - m\Phi$. Indeed, for $\kappa/m \gg \hat{\mu}$ we have
\be
\phi'(r) \simeq  \sqrt{\kappa(r)} \,.
\label{phiMOND}
\ee
In this limit the scalar acceleration on an ordinary matter particle is
\be
a_\phi(r) = \alpha \frac{\Lambda}{M_{\rm Pl}} \phi' \simeq \sqrt{\frac{\alpha^3\Lambda^2}{M_{\rm Pl}}\frac{G_{\rm N} M_{\rm b}(r)}{r^2} }\,.
\label{aphi}
\ee
To reproduce the MONDian result $a_{\rm MOND}  = \sqrt{a_0 \frac{G_{\rm N} M_{\rm b}(r)}{r^2}}$, we are therefore led to identify
\be
\alpha^{3/2} \Lambda = \sqrt{a_0 M_{\rm Pl}}\simeq 0.8~{\rm meV} ~~\Longrightarrow ~~\alpha \simeq 0.86 \left( \frac{\Lambda}{{\rm meV}}\right)^{-2/3}\,,
\label{alphasoln}
\ee
which fixes $\alpha$ in terms of $\Lambda$ through the critical acceleration. For the fiducial value $\Lambda = 0.2$~meV, we obtain $\alpha\simeq 2.5$. 

As it stands, however, the $X < 0$ solution is unstable. It leads to unphysical halos, with growing DM density profiles~\cite{Berezhiani:2015pia,Berezhiani:2015bqa}.
The instability can be seen by expanding Eq.~(\ref{PMOND}) to quadratic order in phonon perturbations $\varphi = \phi - \bar{\phi}(r)$,
\be
{\cal L}_{\rm quad} = {\rm sign}(\bar{X}) \frac{\Lambda(2m)^{3/2}}{4\sqrt{|\bar{X}|}} \left( \dot{\varphi}^2 - 2\frac{\bar{\phi}'}{m} \varphi'\dot{\varphi} - 2\frac{\varphi'^2}{m}\left(\bar{X} - \frac{\bar{\phi}'^2}{2m}\right) - \frac{2\bar{X}}{mr^2} (\partial_\Omega\varphi)^2 \right)\,.
\label{LMONDquad}
\ee
The kinetic term $\dot{\varphi}^2$ has the wrong sign for $\bar{X} < 0$. (The $X > 0$ branch, meanwhile, is stable but does not admit a MOND regime~\cite{Berezhiani:2015pia,Berezhiani:2015bqa}.) 

Since the DM condensate in actual galactic halos has non-zero temperature, however, we expect that the zero-temperature Lagrangian (Eq.~(\ref{PMOND})) to
receive finite-temperature corrections in galaxies. At finite sub-critical temperature, the system is described phenomenologically by Landau's two-fluid model: an admixture of a superfluid component and a normal component. The finite-temperature effective Lagrangian is a function of three scalars~\cite{Nicolis:2011cs}: ${\cal L}_{T \neq 0} = F(X,B,Y)$. The
scalar $X$, already defined in Eq.~(\ref{PMOND}), describes the phonon excitations. The remaining scalars
are defined in terms of the three Lagrangian coordinates $\psi^I(\vec{x},t)$, $I = 1,2,3$ of the normal fluid:
\bea
\nonumber
B &\equiv& \sqrt{{\rm det}\;\partial_\mu\psi^I\partial^\mu\psi^J} \;; \\
Y &\equiv&  u^\mu\left(\partial_\mu\theta + m\delta_\mu^{\;0}\right) -m \simeq  \mu - m\Phi +  \dot{\phi} + \vec{v}\cdot \vec{\nabla}\phi \,,
\label{BY}
\eea
where $u^\mu = \frac{1}{6\sqrt{B}} \epsilon^{\mu\alpha\beta\gamma}\epsilon_{IJK}\partial_\alpha\psi^I\partial_\beta\psi^J \partial_\gamma\psi^K$
is the unit 4-velocity vector, and in the last step for $Y$ we have taken the non-relativistic limit $u^\mu \simeq (1-\Phi, \vec{v})$. By construction, these scalars respect the internal symmetries: $i)$ $\psi^I \rightarrow \psi^I + c^I$ (translations); $ii)$ $\psi^I  \rightarrow R^I_{\;J} \psi^J$ (rotations); $iii)$ $\psi^I \rightarrow \xi^I(\psi)$,
with $\det \frac{\partial\xi^I}{\partial\psi^J} = 1$ (volume-preserving reparametrizations).

There is much freedom in specifying finite-temperature operators that stabilize the MOND profile. The simplest possibility is to supplement~Eq.~(\ref{PMOND}) with the
two-derivative operator
\be
\Delta {\cal L} = M^2Y^2 = M^2(\hat{\mu} +  \dot{\phi})^2\,, 
\ee
where we have specialized to the rest frame of the normal fluid, $\vec{v} = 0$.
This leaves the static profile given by Eq.~(\ref{phigensoln}) unchanged, but modifies the quadratic Lagrangian by $M^2 \dot{\varphi}^2$,
restoring stability for sufficiently large $M$. Specifically this is the case for 
\be
M ~\lower .75ex \hbox{$\sim$} \llap{\raise .27ex \hbox{$>$}}~ \frac{\Lambda m^{3/2}}{\sqrt{|\bar{X}|}} \sim 0.5~\left(\frac{10^{11}\,M_\odot}{M_{\rm b}}\right)^{1/4}  \left(\frac{\Lambda}{{\rm meV}}\right)^{1/2}  \left(\frac{r}{10~{\rm kpc}}\right)^{1/2} \;m \,,
\label{Mbound}
\ee
which, remarkably, is of order eV! Hence, for quite natural values of $M$, this two-derivative operator can restore stability. 
Furthermore, this operator gives a contribution $\Delta P = M^2 \mu^2$ to the condensate pressure, which obliterates the unwanted
growth in the DM density profile. Instead, the pressure is positive far from the baryons, resulting
in localized, finite-mass halos~\cite{Berezhiani:2015pia,Berezhiani:2015bqa}.

\section{Observational Implications} We conclude with some astrophysical implications of our DM superfluid.\\

\noindent {\it Gravitational Lensing:} In TeVeS~\cite{Bekenstein:2004ne} the complete absence of DM requires introducing a time-like vector field $A_\mu$,
as well as a complicated coupling between $\phi$, $A_\mu$ and baryons in order to reproduce lensing observations. In our case, there is no need to introduce an extra vector,
as the normal fluid already provides a time-like vector $u^\mu$. Moreover, our DM contributes to lensing, so we are free to generalize the TeVeS coupling~\cite{Berezhiani:2015pia,Berezhiani:2015bqa}.\\

\noindent {\it Vortices}: When spun faster than a critical velocity, a superfluid develops vortices. The typical angular velocity of halos is well
above critical~\cite{Berezhiani:2015pia,Berezhiani:2015bqa}, giving rise to an array of DM vortices permeating the disc~\cite{Silverman:2002qx}. It will be interesting to see whether these vortices
can be detected through substructure lensing, {\it e.g.}, with ALMA~\cite{Hezaveh:2012ai}.\\

\noindent {\it Galaxy mergers}: A key difference with $\Lambda$CDM is the merger rate of galaxies. Applying Landau's criterion, we find two possible outcomes.
If the infall velocity $v_{\rm inf}$ is less than the phonon sound speed $c_s$ (of order the viral velocity~\cite{Berezhiani:2015pia,Berezhiani:2015bqa}),
then halos will pass through each other with negligible dissipation, resulting in multiple encounters and a longer merger time. If $v_{\rm inf} \;~\lower .75ex \hbox{$\sim$} \llap{\raise .27ex \hbox{$>$}}~\; c_{\rm s}$, however, the
encounter will excite DM particles out of the condensate, resulting in dynamical friction and rapid merger.   \\

\noindent {\it Bullet Cluster}: For merging galaxy clusters, the outcome also depends on the relative fraction of superfluid vs normal components in the clusters.
For subsonic mergers, the superfluid cores should pass through each other with negligible friction (consistent with the Bullet Cluster),
while the normal components should be slowed down by self interactions. Remarkably this picture is consistent with the lensing map of the Abell 520 
``train wreck"~\cite{Mahdavi:2007yp,Jee:2012sr,Clowe:2012am,Jee:2014hja}, which show lensing peaks coincident with galaxies (superfluid components),
as well as peaks coincident with the X-ray luminosity peaks (normal components).\\

\noindent {\it Dark-bright solitons:} Galaxies in the process of merging should exhibit interference patterns (so-called dark-bright solitons) that have been observed in BECs counterflowing at super-critical velocities~\cite{exptpaper}. This can potentially offer an alternative mechanism to generate the spectacular shells seen around elliptical galaxies~\cite{shells}.\\

\noindent {\it Globular clusters:} Globular clusters are well-known to contain negligible amount of DM, and as such pose a problem for MOND~\cite{Ibata:2011ri}. In our case the presence of a significant DM component is necessary for MOND. If whatever mechanism responsible for DM removal in $\Lambda$CDM is also effective here, our model would predict DM-free (and hence MOND-free) globular clusters.

\section*{Acknowledgments}

I wish to thank Lasha Berezhiani for a stimulating collaboration. I also want to warmly thank the organizers and participants of the 2015 Rencontres de Moriond for an inspiring and most enjoyable conference. This work is supported in part by NSF CAREER Award PHY-1145525 and NASA ATP grant NNX11AI95G.


\section*{References}

\begin{thebibliography}{99}

\bibitem{Ade:2013zuv} 
  P.~A.~R.~Ade {\it et al.}  [Planck Collaboration],
  Astron.\ Astrophys.\  {\bf 571}, A16 (2014).

\bibitem{McGaugh:2000sr} 
  S.~S.~McGaugh, J.~M.~Schombert, G.~D.~Bothun and W.~J.~G.~de Blok,
  Astrophys.\ J.\  {\bf 533}, L99 (2000).

\bibitem{McGaugh:2005qe} 
  S.~S.~McGaugh,
  Astrophys.\ J.\  {\bf 632}, 859 (2005).

\bibitem{Vogelsberger:2014dza} 
  M.~Vogelsberger {\it et al.},
  MNRAS\  {\bf 444}, 1518 (2014).

\bibitem{BoylanKolchin:2011de} 
  M.~Boylan-Kolchin, J.~S.~Bullock and M.~Kaplinghat,
  MNRAS\  {\bf 415}, L40 (2011).

\bibitem{Pawlowski:2012vz} 
  M.~S.~Pawlowski, J.~Pflamm-Altenburg and P.~Kroupa,
  MNRAS\  {\bf 423}, 1109 (2012).

\bibitem{Ibata:2013rh} 
  R.~A.~Ibata {\it et al.},
  Nature {\bf 493}, 62 (2013).

\bibitem{Ibata:2014pja} 
  R.~A.~Ibata {\it et al.},
  Astrophys.\ J.\  {\bf 784}, L6 (2014).

\bibitem{Ibata:2014csa} 
  N.~G.~Ibata, R.~A.~Ibata, B.~Famaey and G.~F.~Lewis,
  Nature {\bf 511}, 563 (2014).

\bibitem{Zhao:2013uya} 
  H.~Zhao, B.~Famaey, F.~LŸghausen and P.~Kroupa,
  Astron.\ Astrophys.\  {\bf 557}, L3 (2013).

\bibitem{Milgrom:1983ca} 
  M.~Milgrom,
  Astrophys.\ J.\  {\bf 270}, 365 (1983).

\bibitem{Sanders:2002pf} 
  R.~H.~Sanders and S.~S.~McGaugh,
  Ann.\ Rev.\ Astron.\ Astrophys.\  {\bf 40}, 263 (2002).

\bibitem{Famaey:2011kh} 
  B.~Famaey and S.~McGaugh,
  Living Rev.\ Rel.\  {\bf 15}, 10 (2012).

\bibitem{Broeils1992}
A.~H.~Broeils,
Astron.\ and\ Astrophys. {\bf 256},19 (1992).

\bibitem{Begeman1991}
K.~G.~Begeman, A.~H.~Broeils and R.~H.~Sanders,
MNRAS\  {\bf 249}, 523 (1991).

\bibitem{sellwood2005}
J.~A.~Sellwood and S.~S.~McGaugh,
Astrophys.\ J.\  {\bf 634}, 70 (2005).

\bibitem{Milgrom2009}
  M.~Milgrom,
  arXiv:0908.3842 [astro-ph.CO].

\bibitem{Aguirre:2001fj} 
  A.~Aguirre, J.~Schaye and E.~Quataert,
  Astrophys.\ J.\  {\bf 561}, 550 (2001).

\bibitem{Bekenstein:2004ne} 
  J.~D.~Bekenstein,
  Phys.\ Rev.\ D {\bf 70}, 083509 (2004)
  [Erratum-ibid.\ D {\bf 71}, 069901 (2005)].

\bibitem{Skordis:2005xk} 
  C.~Skordis, D.~F.~Mota, P.~G.~Ferreira and C.~Boehm,
  Phys.\ Rev.\ Lett.\  {\bf 96}, 011301 (2006).

\bibitem{Clowe:2003tk} 
  D.~Clowe, A.~Gonzalez and M.~Markevitch,
  Astrophys.\ J.\  {\bf 604}, 596 (2004).

\bibitem{Clowe:2006eq} 
  D.~Clowe {\it et al.},
  Astrophys.\ J.\  {\bf 648}, L109 (2006).

\bibitem{Angus:2006qy} 
  G.~W.~Angus, B.~Famaey and H.~Zhao,
  MNRAS\  {\bf 371}, 138 (2006).

\bibitem{Berezhiani:2015pia} 
  L.~Berezhiani and J.~Khoury,
  arXiv:1506.07877 [astro-ph.CO].

\bibitem{Berezhiani:2015bqa} 
  L.~Berezhiani and J.~Khoury,
  arXiv:1507.01019 [astro-ph.CO].

\bibitem{tisza}
L.~Tisza,
C.\ R.\ Acad.\ Sci.\ {\bf 207}, 1035 (1938); {\bf 207}, 1186 (1938).

\bibitem{london}
F.~London, 
Phys.\ Rev.\  {\bf 54}, 947 (1938).

\bibitem{landau}
L.~D. ~Landau, 
J. Phys. (USSR) {\bf 5}, 71 (1941); {\bf 11}, 91 (1947).

\bibitem{Harvey:2015hha} 
  D.~Harvey {\it et al.},
  Science {\bf 347}, no. 6229, 1462 (2015).

\bibitem{Son:2002zn} 
  D.~T.~Son,
  hep-ph/0204199.

\bibitem{UFGreview}
W.~Zwerger, ed.
{\it The BCS-BEC Crossover and the Unitary Fermi Gas},
Lecture Notes in Physics, Vol. 836 (Springer- Verlag, Berlin Heidelberg, 2012).

\bibitem{Giorgini:2008zz} 
  S.~Giorgini, L.~P.~Pitaevskii and S.~Stringari,
  Rev.\ Mod.\ Phys.\  {\bf 80}, 1215 (2008).

\bibitem{Son:2005rv} 
  D.~T.~Son and M.~Wingate,
  Annals Phys.\  {\bf 321}, 197 (2006).

\bibitem{Villain}
  J.~Villain,
  J.\ Phys.\ (France) {\bf 36}, 581 (1975).

\bibitem{cosinecoupling}
J.~V.~Jos\'e, L.~P.~Kadanoff, S.~Fitzpatrick and D.~R.~Nelson,
Phys.\ Rev.\  B {\bf 16}, 1217 (1977).

\bibitem{Randy}
  R.~D.~Kamien,
  HUTP-89/A025.

\bibitem{chandrabook}
S.~Chandrasekhar,
``An introduction to the study of stellar structure,"
Dover Publications, New York (1957).

\bibitem{Nicolis:2011cs} 
  A.~Nicolis,
  arXiv:1108.2513 [hep-th].

\bibitem{Silverman:2002qx} 
  M.~P.~Silverman and R.~L.~Mallett,
  Gen.\ Rel.\ Grav.\  {\bf 34}, 633 (2002).

\bibitem{Hezaveh:2012ai} 
  Y.~Hezavehv{\it et al.},
  Astrophys.\ J.\  {\bf 767}, 9 (2013).
 
\bibitem{Mahdavi:2007yp} 
  A.~Mahdavi {\it et al.},
  Astrophys.\ J.\  {\bf 668}, 806 (2007).

\bibitem{Jee:2012sr} 
  M.~J.~Jee {\it et al.},
  Astrophys.\ J.\  {\bf 747}, 96 (2012).

\bibitem{Clowe:2012am} 
  D.~Clowe {\it et al.},
  Astrophys.\ J.\  {\bf 758}, 128 (2012).

\bibitem{Jee:2014hja} 
  M.~J.~Jee {\it et al.},
  Astrophys.\ J.\  {\bf 783}, 78 (2014).

\bibitem{exptpaper}
C.~Hamner, J.~J.~Chang, P.~Engels and M.~A.~Hoefer,
Phys.\ Rev.\ Lett.\  {\bf 106}, 065302 (2011).

\bibitem{shells}
A.~P.~Cooper {\it et al.}, 
Astrophys. J. Lett. {\bf 743}, L21 (2011).

\bibitem{Ibata:2011ri} 
  R.~Ibata {\it et al.},
  Astrophys.\ J.\  {\bf 738}, 186 (2011).




\end{thebibliography}
\end{document}